# Exceptional Topological Bound States in the Continuum

Luohong Liu, Qicheng Zhang[†], Shuaishuai Tong, and Chunyin Qiu[*]

Key Laboratory of Artificial Micro- and Nano-Structures of Ministry of Education
and School of Physics and Technology, Wuhan University, Wuhan 430072, China

[†]Contact author: qicheng.zhang@whu.edu.cn

[*]Contact author: cyqiu@whu.edu.cn

*Abstract.* Topological phases, exceptional points, and bound states in the continuum (BICs) are three key concepts in condensed matter and classical wave physics. Their pairwise combinations have yielded a rich variety of intriguing singular modes, but the triple merger—exceptional topological BICs (ET-BICs)—has yet to be explored. In this Letter, we theoretically propose and experimentally demonstrate a universal approach for constructing such elusive exceptional structures. The system consists of four coupled topological chains: $y$-mirror stacking generates pairs of topological BICs, while $z$-stacking with biased loss yields a generalized mirror symmetry under which each pair—separated in different subspaces until their collapse—merges into an ET-BIC at a critical loss. Experimentally, using an acoustic platform with electrically tunable loss, we not only directly observe the coalescence of eigenenergies but also the coalescence of eigenstates, offering a complete verification of the ET-BIC formation. The triply synergistic ET-BICs may pave the way for wave devices that simultaneously demand robustness, sensitivity, and BIC confinement.

*Introduction*. Topological materials have attracted intense interest across condensed matter physics and beyond [1-6]. A significant characteristic of topological phases is the bulk-boundary correspondence, which ties nontrivial bulk topology to robust localized modes within band gaps [1,2]. In contrast, bound states in the continuum (BICs) are localized modes embedded inside the extended continuum, seemingly circumventing the gap constraint [7,8]. However, BICs are fragile and easily become leaky resonances against perturbations. The natural combination of these two concepts—topological BICs (T-BICs)—extends the bulk-boundary correspondence paradigm beyond the band gap while stabilizing the BICs with topological protection [9-25]. When extended to non-Hermitian settings, qualitatively new phenomena emerge. Exceptional points (EPs) [26-32] stand out as intrinsic non-Hermitian singularities, where the simultaneous coalescence of eigenvalues and eigenvectors yields enhanced spectral sensitivity. Pairing EPs with topological edge modes (TEMs) gives rise to topological EPs (T-EPs), uniting topological robustness with EP-enhanced sensitivity [33-41]; while coupling EPs with BICs yields exceptional BICs (E-BICs), merging singular spectral response with non-radiative confinement [42-45]. These pairwise synergies demonstrate the power of concept fusion and, naturally, motivate a simultaneous three-way combination (see Fig. 1).

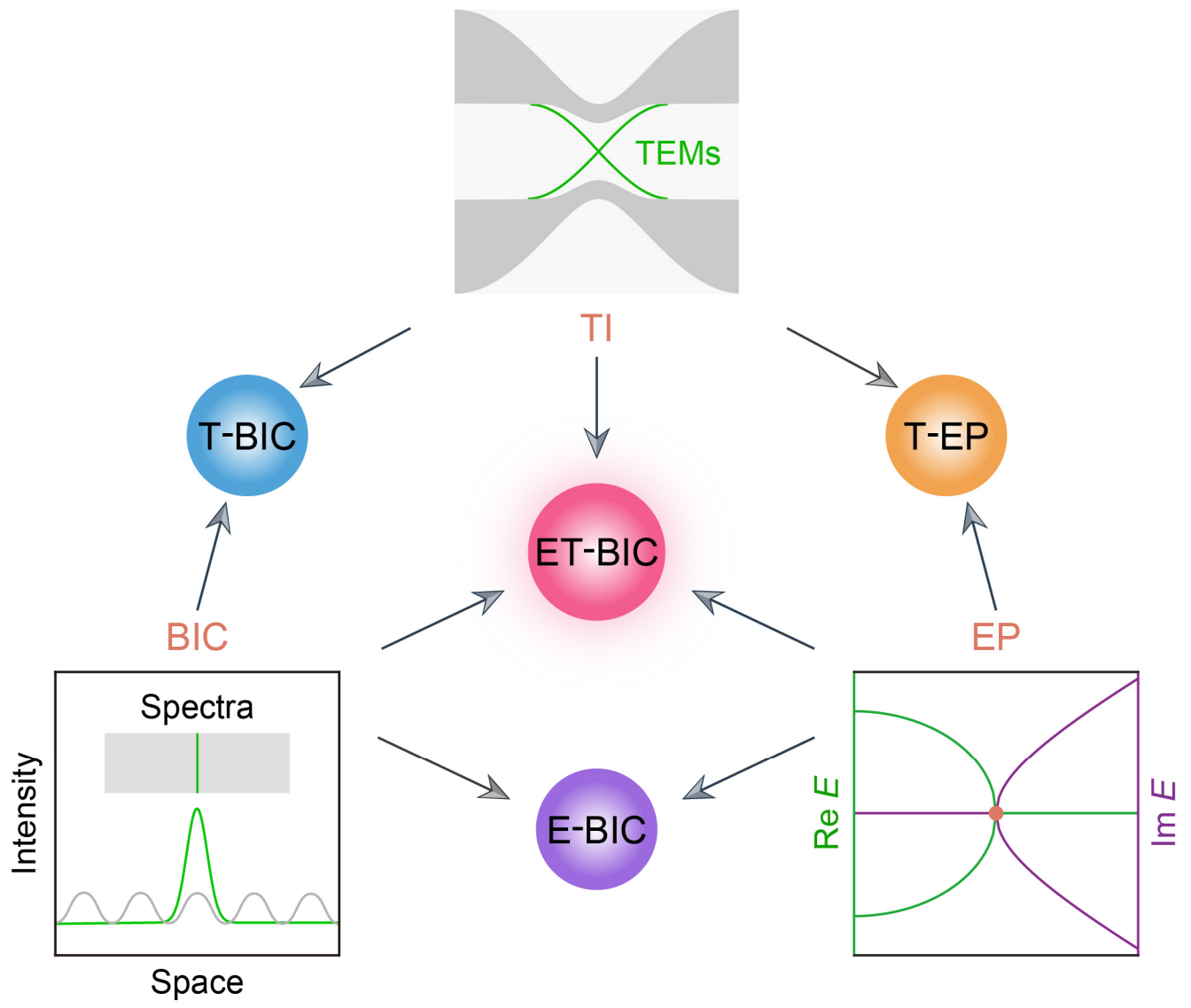

FIG. 1. Conceptual framework of ET-BICs emerging from topology, BICs, and EPs. Arrows trace the progression from the three ingredients to their pairwise combinations (T-BICs, T-EPs, E-BICs), and ultimately to the central ET-BICs that bring together topological protection, BIC confinement, and EP sensitivity.

To the best of our knowledge, the triple merger of topology, BICs, and EPs—exceptional topological BICs (ET-BICs), a singular structure that unlocks functionalities beyond any pairwise

platform—has yet to be explored theoretically, let alone demonstrated experimentally. Nevertheless, realizing such exceptional structures poses substantial challenges on both theory and experiment. Theoretically, one must establish a unified framework that simultaneously accommodates the three key ingredients, i.e., topological protection, BIC confinement, and EP formation, and that, with a single parameter, enables the coalescence of BICs while preserving their nontrivial topology. Experimentally, the challenges are multiple and escalating. In addition to constructing a physical platform that hosts all three ingredients, one must distinguish the constituent BICs (without which their coalescence cannot be precisely tracked) and, ultimately, probe both the eigenenergy and eigenstate coalescence for a complete verification of the ET-BIC.

In this Letter, we establish a unified theoretical framework—bringing the three key ingredients together—for constructing ET-BICs. As illustrated in Fig. 2, our system consists of four $x$-directed topological chains, $y$-mirror-stacked into two chain pairs and further $z$-stacked with biased loss. The $y$-mirror-stacking yields two pairs of energetically separated, $y$-parity-locked T-BICs, while the $z$-stacking exhibits a generalized mirror symmetry (GMS) that preserves a clean subspace decomposition and enables selective access to the T-BICs of specific GMS parity. As we increase the biased loss towards a critical value, the two GMS subspaces coalesce and each ($y$-parity locked) T-BIC pair merges into an ET-BIC; beyond this point, the GMS subspaces revive and each ET-BIC re-splits into two T-BICs of opposite GMS parity. Experimentally, we implement this approach in an acoustic platform with continuously tunable loss, and directly observe the spectral evolution by tracking the frequencies and linewidths of resonance peaks, while characterizing the eigenstates' coalescence via probing their parities and inner products. Our construction is universal for any elementary topological chain, and its complete experimental demonstration establishes a general route to exceptional structures with coexisting topological protection, BIC confinement, and EP sensitivity.

*GMS-enabled ET-BICs via subspace coalescence*. We start from a generic topological chain $h_0$ that hosts in-gap TEMs [Fig. 2(a)]. By mirror-stacking $h_0$ along $y$ [Fig. 2(b)], we build a chain pair with Hamiltonian

$$H_0 = \sigma_0 \otimes h_0 + t_y \sigma_1 \otimes I_{h_0}, \tag{1}$$

where $\sigma_i$ are Pauli matrices, $t_y$ represents the inter-chain hopping along $y$, and $I_{h_0}$ is the identity matrix with the same dimension as $h_0$. (For simplicity, here we assume that the inter-chain coupling is uniform across the internal degrees of freedom of $h_0$). The chain pair exhibits conventional mirror symmetry $M_y$, which decomposes it into two subspaces of opposite $y$-mirror parity. Each subspace inherits the spectrum of $h_0$, shifted by $\pm t_y$. By tuning $t_y$, the in-gap TEMs in one subspace can move spectrally into the bulk continuum of the other subspace without hybridization [25], which yields a pair of T-BICs of opposite $y$-parity within an appropriate range of $t_y$.

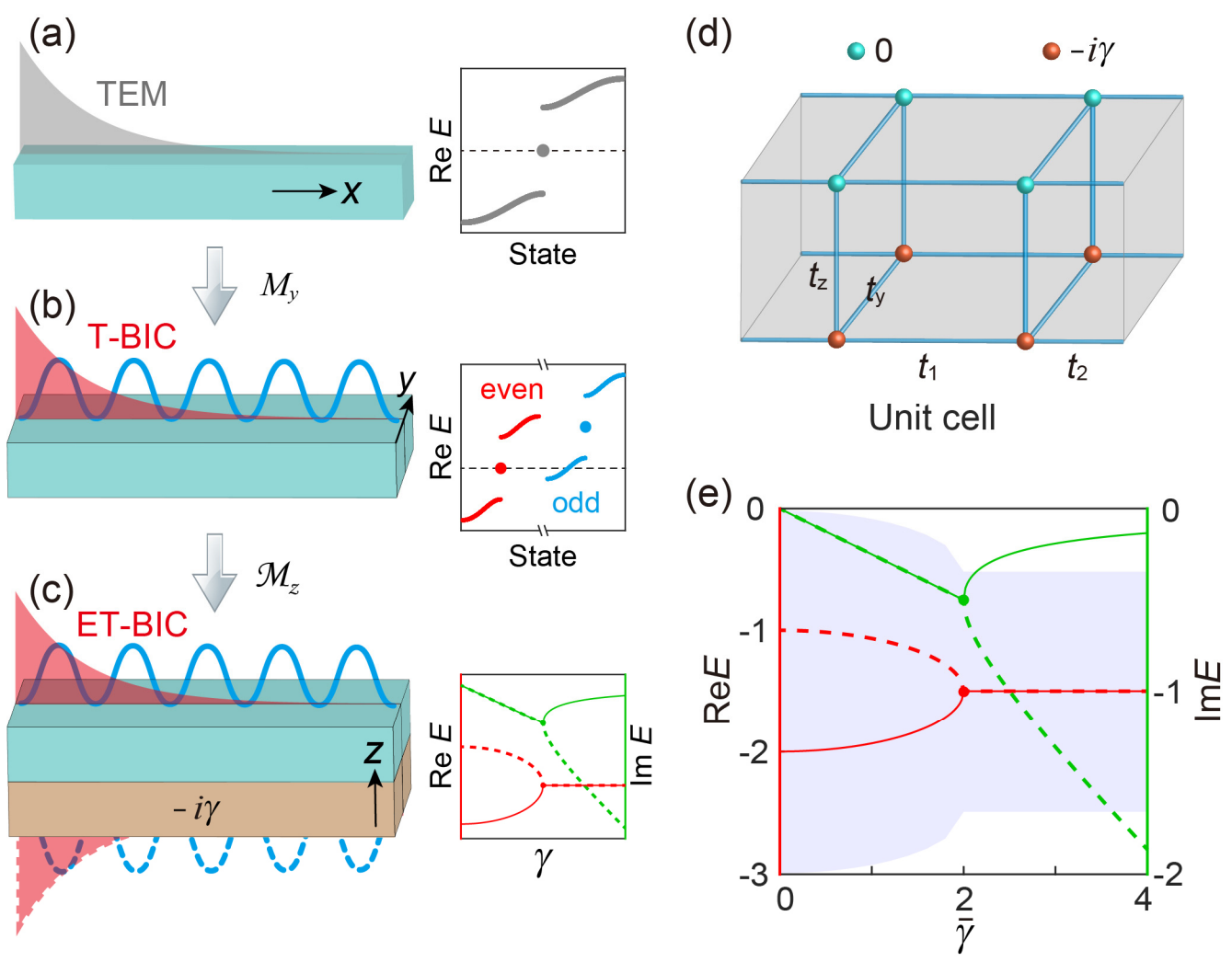

FIG. 2. Layer-stacking construction of ET-BICs via GMS-enabled subspace coalescence. (a)-(c) Schematic illustration of the general procedure: starting from a single topological chain with spectrally isolated TEMs (a), constructing T-BIC pairs via conventional $y$-mirror stacking (b), and realizing ET-BICs via generalized $z$-mirror stacking with biased loss $-i\gamma$ (c). Left panels: structural configurations; right panels: corresponding eigenenergy spectra. (d) Minimal lattice realization: four coupled SSH chains with layer-biased loss $-i\gamma$. (e) Energy spectrum evolution versus the dimensionless loss $\bar{\gamma} = \gamma/t_z$, showing the formation of an ET-BIC in the negative real-energy region. The T-BICs (lines) belong to the even $y$-mirror subspace and are embedded in the odd bulk continuum (shaded region). For $\bar{\gamma} < 2$, all bulk and T-BIC eigenstates have degenerate imaginary parts; for $\bar{\gamma} > 2$, they split into two branches of opposite GMS parity. The solid (dashed) line represents the T-BIC of even (odd) GMS parity.

Based on the chain pair structure with T-BICs, we further duplicate it along $z$ with uniform interlayer coupling $t_z$ [Fig. 2(c)]. To access non-Hermitian EP physics, we introduce biased onsite loss $-i\gamma$ on the bottom layer, which breaks the conventional mirror symmetry $M_z$. The Hamiltonian of the four-chain system reads

$$H = \tau_0 \otimes \left(H_0 - \frac{i\gamma}{2} I_{H_0}\right) + \frac{i\gamma}{2}\tau_3 \otimes I_{H_0} + t_z \tau_1 \otimes I_{H_0}, \tag{2}$$

with $\tau_i$ Pauli matrices and $I_{H_0}$ the identity matrix with the same dimension as $H_0$. Consider the eigenproblem $H\Psi = E\Psi$, with $\Psi = (\Psi_t, \Psi_b)^T$ comprising the top- and bottom-layer components $\Psi_t$ and $\Psi_b$. The eigenvalues of $H$ satisfy

$$E = E_0 - \frac{i\gamma}{2} \pm \sqrt{t_z^2 - \frac{\gamma^2}{4}}, \tag{3}$$

with $E_0$ being the eigenvalue of $H_0$. The spectrum splits into two distinct branches, each reproducing the $H_0$-spectrum up to an energy offset. This unique spectral structure is reminiscent of the above mirror-symmetric case, where the eigenvalues separate into two groups corresponding to different mirror subspaces—yet here the conventional $M_z$ is already broken. Motivated by this analogy, we posit that the four-chain system exhibits a generalized mirror symmetry (GMS), $\mathcal{M}_z$, under which the eigenstates $\Psi$ satisfy $\Psi_b = \alpha\Psi_t$. Substituting this hypothesis into $H\Psi = E\Psi$ yields $t_z\alpha^2 + i\gamma\alpha - t_z = 0$ with the two solutions

$$\alpha_{1,2} = (-i\bar{\gamma} \pm \sqrt{4 - \bar{\gamma}^2})/2, \tag{4}$$

where the dimensionless quantity $\bar{\gamma} = \gamma/t_z$ characterizes a reduced loss strength. (This further makes the spectrum structure clearer, $E = E_0 + \alpha_{1,2}t_z$; at $\bar{\gamma} = 0$, the GMS reduces to the conventional mirror symmetry with $\alpha_{1,2} = \pm 1$.) Based on the form of $\Psi$ in the layer basis, $(1, \alpha_{1,2})^T$, we construct a (generally non-unitary) similarity transformation matrix

$$S = \begin{pmatrix} 1 & 1 \\ \alpha_1 & \alpha_2 \end{pmatrix} \otimes I_{H_0}, \tag{5}$$

which decouples $H$ into two independent subspaces via $\widetilde{H} = S^{-1}HS = (H_0 + \alpha_1 t_z) \oplus (H_0 + \alpha_2 t_z)$ (see Fig. S1 in Supplemental Material for an illustration). These subspaces carry different GMS parities, with the corresponding operator $\widetilde{\mathcal{M}}_z = S^{-1}\mathcal{M}_z S = I_{H_0} \oplus (-I_{H_0})$. Therefore, the GMS operator in the original basis reads

$$\mathcal{M}_z = S\widetilde{\mathcal{M}}_z S^{-1} = \frac{1}{\sqrt{4-\bar{\gamma}^2}} \begin{bmatrix} i\bar{\gamma} & 2 \\ 2 & -i\bar{\gamma} \end{bmatrix} \otimes I_{H_0}. \tag{6}$$

This explicit form allows us to verify the commutation relation $[H, \mathcal{M}_z] = 0$, thereby establishing the GMS hypothesis as an inherent mathematical structure of the loss-biased system.

Note that the commutation relation $[H, \mathcal{M}_z] = 0$ holds only for $\bar{\gamma} \neq 2$ (equivalently, $\alpha_1 \neq \alpha_2$). At $\alpha_1 = \alpha_2$, the two GMS subspaces collapse into a single exceptional subspace—a feature recently termed an exceptional deficiency, which generalizes the EP concept to the coalescence of entire eigenspaces [46]. Unlike that scheme, where the deficiency is formed by unidirectionally coupling two systems with identical spectra, our mechanism achieves this coalescence via layer-biased loss within a GMS framework. By tuning the biased loss, we can continuously drive the subspaces into and out of the exceptional deficiency. As illustrated in Fig. 2(c), a pair of T-BICs of opposite GMS parity merge into an ET-BIC at the critical loss $\bar{\gamma} = 2$, accompanied by the collapse of GMS subspaces. Crucially, the topology of the system is independent of $\bar{\gamma}$: in the GMS-decomposed form $\widetilde{H} = (H_0 + \alpha_1 t_z) \oplus (H_0 + \alpha_2 t_z)$, the loss enters only through the energy offsets $\alpha_{1,2}t_z$, whereas $H_0$—and hence its topological invariant—remains unchanged. (The divergence of $\mathcal{M}_z$ at $\bar{\gamma} = 2$ reflects the GMS-subspace collapse rather than a topological transition.) Therefore, the topological protection of the TEMs is preserved throughout the ET-BIC formation.

Our GMS-based approach to ET-BICs is universal for any elementary topological chain, as it does not depend on the specific form of $h_0$. As a concrete example, here we consider the Su–Schrieffer–Heeger (SSH) chain for $h_0$, the minimal topological model. The hopping parameters are $t_1 = -1$, $t_2 = -3$, $t_y = -1.5$, and $t_z = -0.5$. Figure 2(e) presents the energy spectrum versus the dimensionless loss $\bar{\gamma} = \gamma/t_z$ (restricted to the negative real-energy region for conciseness). It clearly shows the evolution of a pair of T-BICs of even $y$-mirror parity embedded in the bulk continuum of odd $y$-mirror parity: as $\bar{\gamma}$ increases, the two T-BICs of opposite GMS parity approach each other, and at $\bar{\gamma} = 2$ they coalesce into a single ET-BIC, signaling the collapse of the two GMS subspaces. The complete spectrum is provided in Supplemental Material (Fig. S2), which also exhibits the formation of an ET-BIC from the $y$-odd T-BIC pair in the positive-energy region.

*Experimental observation of ET-BICs in an acoustic platform.* As shown in Fig. 3(a), our sample consists of four $x$-directed cavity-tube chains, forming two $y$-mirror-stacked acoustic SSH chain pairs and further $z$-stacked with layer-biased loss. In this acoustic platform, the air cavities emulate lattice sites and the narrow tubes provide reciprocal couplings between them (see Supplemental Material, Figs. S3-S4). The cavity resonance ($\approx 3950$ Hz) defines the onsite energy that sets the zero-frequency reference, while the tube geometries yield effective hoppings $t_1 \approx -36.8$ Hz, $t_2 \approx -154.8$ Hz, $t_y \approx -76.9$ Hz, and $t_z \approx -27.6$ Hz, which ensure the formation of T-BICs and the subsequent observation of the ET-BIC evolution. Beyond the uniform background loss $-i\gamma_0 \approx -27.0i$ Hz, a controllable loss $-i\gamma$ is introduced into the bottom layer via electrical circuits (see Supplemental Material, Fig. S5). Note that two T-BIC pairs reside in different $y$-mirror subspaces, and within each subspace, the two T-BICs—further distinguished by their GMS parity—undergo independent coalescence into a single ET-BIC. To separately track their pairwise evolutions and precisely identify the eigenstate coalescence, we selectively excite the four mirror subspaces characterized by $y$-parity and GMS parity. To do this, we place four point-like sources on the four chains, all sharing the same $x$ coordinate (either edge or bulk sites). The excitation is controlled at two levels: first, the two chains within each $y$-mirror chain pair are excited in phase or out of phase to select the $y$-even or $y$-odd subspace; second, the relative amplitude and phase between the bottom and top layers are set according to the eigenstate structure $\Psi \sim (1, \alpha_{1,2})^T$ in the layer basis, with $\alpha_1$ for the GMS-even subspace and $\alpha_2$ for the GMS-odd subspace.

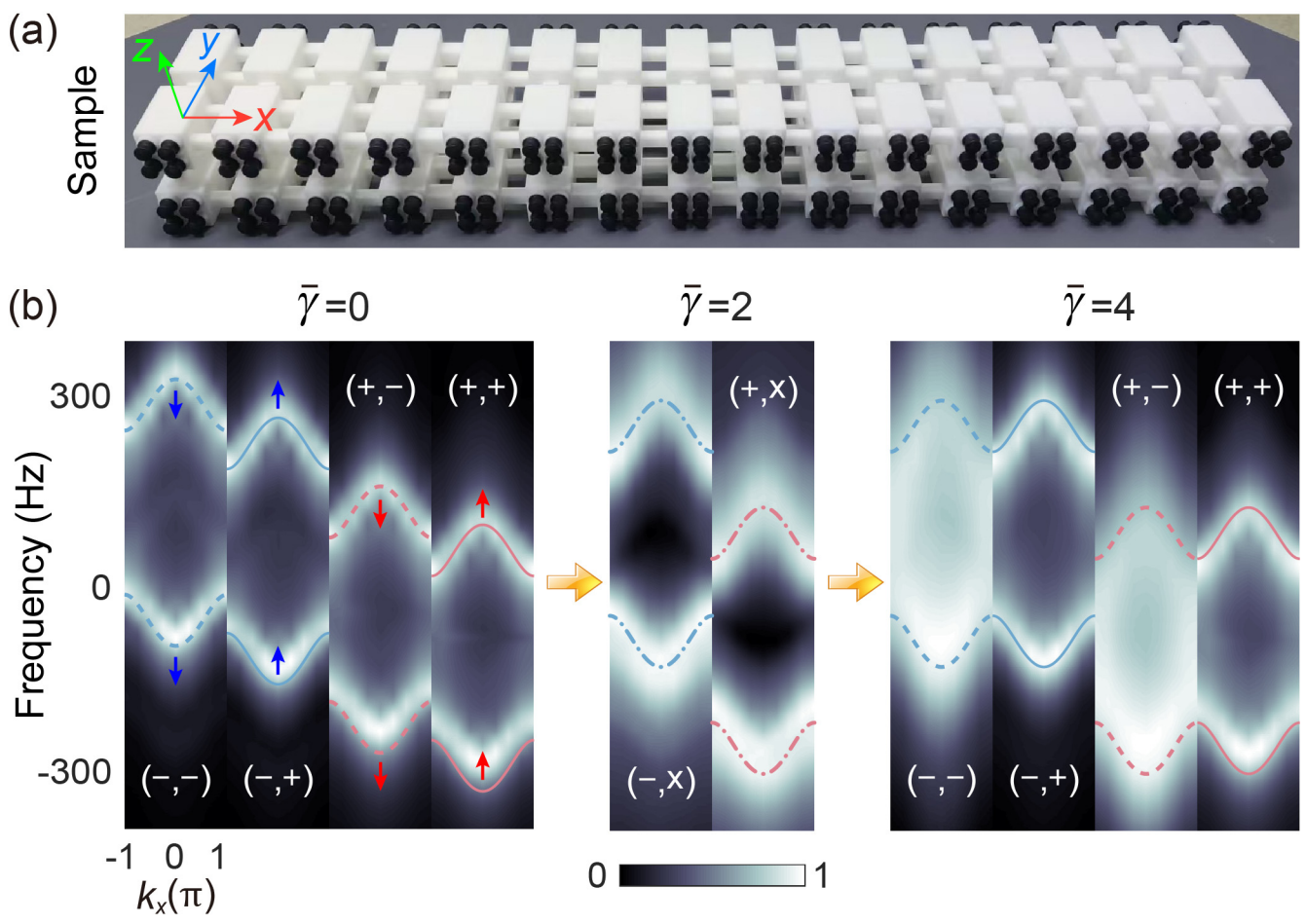

FIG. 3. Momentum-space observation of GMS subspace collapse and revival. (a) Experimental sample composed of four nontrivial acoustic SSH chains, with biased loss on the bottom layer controlled by electrical circuits (not shown). (b) Measured band structures (color scale) for three representative $\bar{\gamma}$ values, overlaid with theoretical band structures (lines). The selectively excited subspaces are labeled by their parities ($M_y$,$\mathcal{M}_z$), where × denotes an undefined GMS parity at $\bar{\gamma} = 2$.

We first measure the momentum-space band structures to reveal how the GMS subspaces evolve with the biased loss [Fig. 3(b)]. The band structures are obtained by placing the sources at the chain center and Fourier transforming the response over the sample. At $\bar{\gamma} = 0$, the four subspaces are clearly resolved. In particular, the bands from different GMS subspaces are centered at different energies while exhibiting identical broadening. These behaviors are consistent with Eq. (3): energy splitting in the real parts, while degeneracy in the imaginary parts (set by the uniform background loss). As $\bar{\gamma}$ increases, the bands belonging to the same $y$-parity but opposite GMS parity move toward each other, as indicated by arrows. At $\bar{\gamma} = 2$ (where the exceptional deficiency occurs), the four subspaces collapse into two exceptional subspaces, while each band-pair of the same $y$-parity coalesces into an exceptional band. At $\bar{\gamma} = 4$, the GMS subspaces revive and each exceptional band re-splits into two bands of opposite GMS parity. In sharp contrast to $\bar{\gamma} = 0$, here the bands of the same $y$-parity yet opposite GMS parity share the same frequencies but acquire markedly different band broadening, as predicted by Eq. (3). These observations clearly demonstrate the collapse and revival of the GMS subspaces, which underlie the T-BIC coalescence into ET-BICs that we examine next.

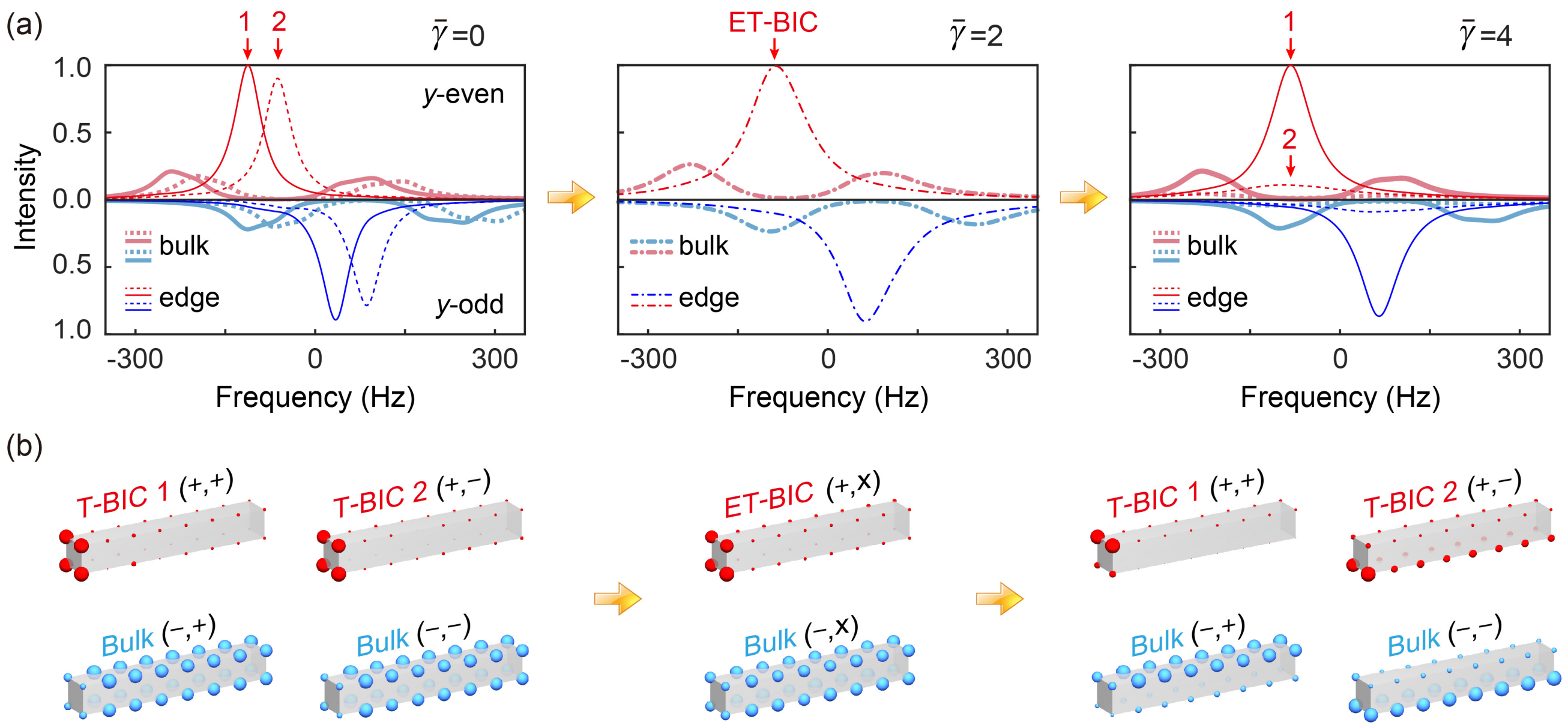

FIG. 4. Spectral and spatial identification of T-BICs and ET-BICs. (a) Site-averaged edge and bulk intensity spectra measured for three representative $\bar{\gamma}$ values. The paired T-BIC peaks merge into ET-BIC peaks at $\bar{\gamma} = 2$ and re-split thereafter. (b) Measured field distributions at the peak frequencies of T-BICs 1 and 2 [indicated by arrows in (a)]. These T-BICs, belonging to the $y$-even subspace, have opposite GMS parity.

To confirm the formation of T-BICs and their pairwise coalescence into ET-BICs, we measure the spectral responses and spatial distributions for three representative $\bar{\gamma}$ values. We place the sources at the sample edge or in the bulk according to the desired parities ($M_y$,$\mathcal{M}_z$), and detect the corresponding site-resolved, broadband local sound responses. Figure 4(a) presents the site-averaged edge and bulk intensity spectra for the $y$-even and $y$-odd subspaces, and Fig. 4(b) displays the field distributions at specified peak frequencies for each $\bar{\gamma}$. At $\bar{\gamma} = 0$, for each $y$-mirror subspace, two sharp edge peaks coincide in frequency with the broad bulk peaks of opposite $y$-parity, indicating a typical spectral signature of T-BICs. (For clarity, we label the $y$-even T-BICs as 1 and 2, and the $y$-odd ones as 3 and 4.) This identification is further confirmed by the field distributions at the peak frequencies of T-BICs 1 and 2, where the localized $y$-even TEMs coexist with the $y$-odd extended bulk states; similar behaviors are observed for T-BICs 3 and 4 (not shown). At $\bar{\gamma} = 2$, the two T-BICs within each $y$-mirror subspace coalesce into a single ET-BIC peak with broader linewidth. The corresponding ET-BIC field distribution remains edge-localized, consistent with its origin from the coalescence of the two T-BICs. At $\bar{\gamma} = 4$, each ET-BIC splits into two T-BIC peaks that exhibit the same center frequency but contrasting linewidths—in contrast to $\bar{\gamma} = 0$, where the paired T-BICs have identical linewidths but different center frequencies—as predicted in Fig. 2(e). The re-split T-BICs, while remaining edge-localized, become layer-contrasting in their intensity distributions: T-BICs 1 and 2—both with layer-independent intensity distributions before the coalescence—are now dominated by the top and bottom layers, respectively, as predicted by their eigenstates in the layer basis. Together, all the spectra and field distributions directly confirm the T-BIC and ET-BIC identities at each stage.

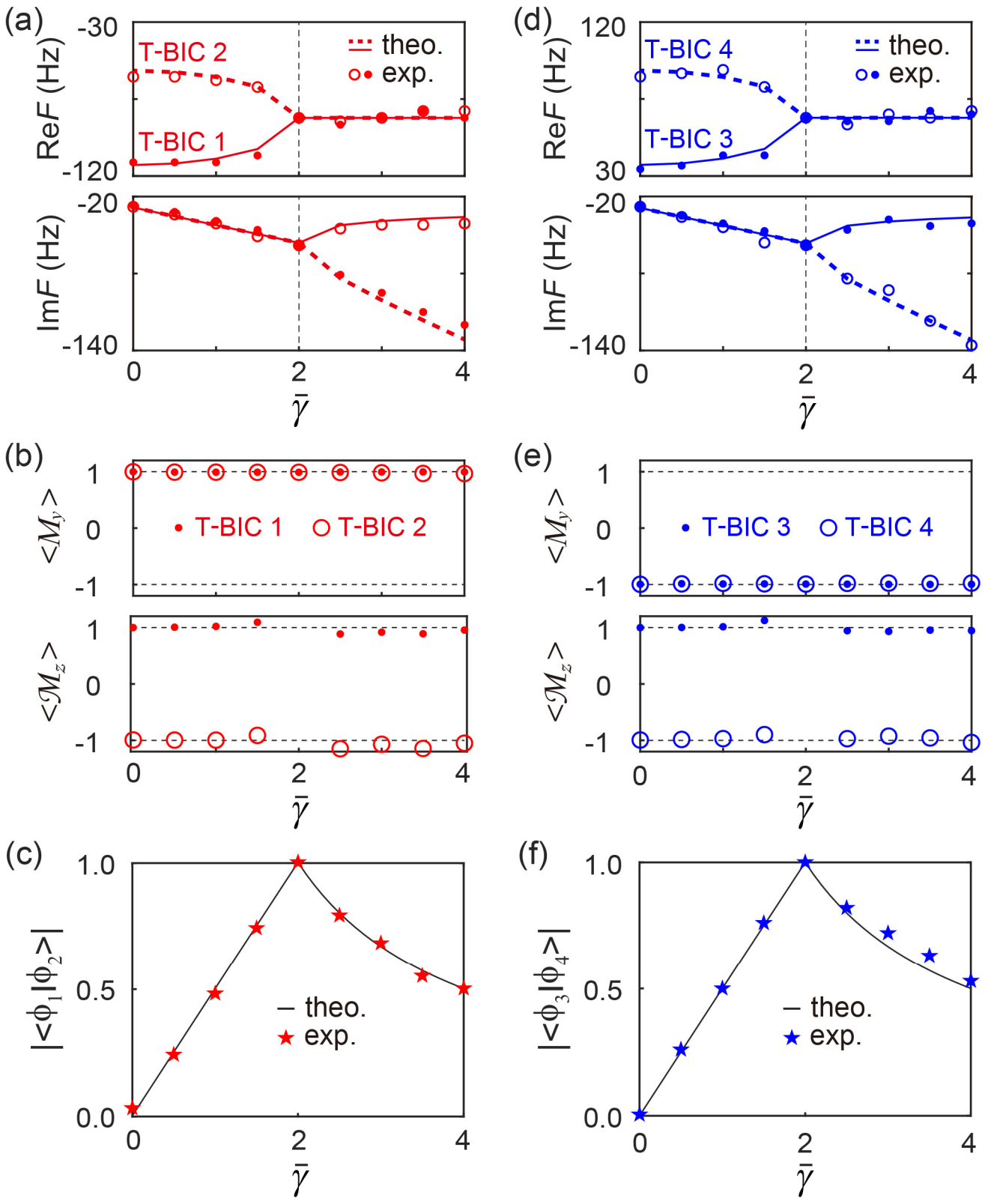

FIG. 5. Experimental tracking of eigenenergy and eigenstate coalescence. (a) Real and imaginary parts of T-BICs 1 and 2 (with $y$-even parity) extracted from a series of $\bar{\gamma}$-dependent T-BIC spectra, matching well with theoretical predictions. (b) Evolution of $y$-parity and GMS parity for these two T-BICs. The GMS parity is undefined at $\bar{\gamma} = 2$. (c) Modulus of the inner product between T-BICs 1 and 2, which reaches 1 at $\bar{\gamma} = 2$ and signifies the formation of an ET-BIC. (d)-(f) Similar to (a)-(c), but for T-BICs 3 and 4 with $y$-odd parity.

We now provide a quantitative characterization of the coalescence process from both the eigenenergy and eigenstate perspectives. To this end, we measure the edge spectral responses and the field distributions for a series of $\bar{\gamma}$ values that span the ET-BIC transition (where the GMS subspaces collapse). Below we focus on the $y$-even subspace [Figs. 5(a)-(c)], with the $y$-odd subspace behaving similarly [Figs. 5(d)-(f)]. Figure 5(a) presents the real and imaginary parts of the eigenenergies for T-BICs 1 and 2, extracted from the peak frequencies and linewidths of the edge spectral responses (see Supplemental Material, Fig. S6). Below $\bar{\gamma} = 2$, the real parts are separated while the imaginary parts are degenerate; above $\bar{\gamma} = 2$, the roles are reversed—the former become degenerate while the latter split. At $\bar{\gamma} = 2$, both the real and imaginary parts coincide, a hallmark of the eigenenergy coalescence. Next, we examine the evolution of eigenstates—the other indispensable ingredient for establishing ET-BICs. Figure 5(b) shows the expectation values of the mirror operators $\langle M_y \rangle$ and $\langle \mathcal{M}_z \rangle$ for T-BICs 1 and 2. Throughout the entire range, $\langle M_y \rangle$ remain close to $+1$ and $\langle \mathcal{M}_z \rangle$ approaches $\pm 1$ (at $\bar{\gamma} \neq 2$), confirming that the two T-BICs are indeed excited in the correct subspaces. Notably, the quantized $\langle \mathcal{M}_z \rangle$ becomes undefined at $\bar{\gamma} = 2$ (blank in the figure) and signals, once again, the collapse of GMS subspaces. To further verify whether the two T-BICs coincide, Fig. 5(c) provides the inner product between their field distributions. It shows that at $\bar{\gamma} = 2$, the inner product increases to 1 and the two T-BICs become indistinguishable—the other defining feature of the ET-BIC, in addition to the spectral coalescence displayed in Fig. 5(a). Together, all the three measurements completely characterize the coalescence of the T-BICs into ET-BICs.

*Conclusion*. We have theoretically proposed and experimentally demonstrated a universal mechanism for constructing ET-BICs by continuously tuning the biased loss within a GMS-based framework. In an acoustic platform, we directly observed the full coalescence and re-splitting of two T-BIC pairs into and out of two ET-BICs, from not only the eigenenergy but also the eigenstate perspective, in tandem with the collapse and revival of GMS subspaces. This mechanism does not rely on any specific Hamiltonian form of the elementary topological chain and can be applied to diverse platforms such as photonic crystals and topological circuits. More generally, the GMS-stacking operates at the level of T-BIC pairs, independent of how the T-BICs are generated. This opens the possibility of extending the construction to two-dimensional topological systems, where paired first-order or higher-order T-BICs may likewise be driven to coalesce. Moreover, our work identifies the GMS as an inherent mathematical structure that underpins the subspace coalescence and the subsequent formation of exceptional structures such as ET-BICs. This symmetry also yields a clean subspace decomposition that enables independent tracking of the two T-BICs in our experiments. Our findings establish a GMS-based design principle for architecting novel non-Hermitian exceptional structures, and offer a route to robust and sensitive wave devices.

**Data availability.** The data that support the findings of this study are available from the corresponding authors upon reasonable request.

**Acknowledgements.** This work was supported by the National Key R&D Program of China (Grant No. 2023YFA1406900), the National Natural Science Foundation of China (Grants No. 12374418, No. 12674584, and No. 12304495), and the Natural Science Foundation of Hubei Province of China (No. 2024AFB654).